# Biaxial versus uniaxial strain tuning of single-layer MoS$_2$

*Felix Carrascoso[1], Riccardo Frisenda[1] (\*), Andres Castellanos-Gomez[1] (\*)*

[1]*Materials Science Factory. Instituto de Ciencia de Materiales de Madrid (ICMM-CSIC), 28049, Madrid, Spain.*

Riccardo.frisenda@csic.es , Andres.castellanos@csic.es

ABSTRACT

Strain engineering has arisen as a powerful technique to tune the electronic and optical properties of two-dimensional semiconductors like molybdenum disulfide (MoS$_2$). Although several theoretical works predicted that biaxial strain would be more effective than uniaxial strain to tune the band structure of MoS$_2$, a direct experimental verification is still missing in the literature. Here we implemented a simple experimental setup that allows to apply biaxial strain through the bending of a cruciform polymer substrate. We used the setup to study the effect of biaxial strain on the differential reflectance spectra of 12 single-layer MoS$_2$ flakes finding a redshift of the excitonic features at a rate between -40 meV/% and -110 meV/% of biaxial tension. We also directly compare the effect of biaxial and uniaxial strain on the same single-layer MoS$_2$ finding that the biaxial strain gauge factor is 2.3 times larger than the uniaxial strain one.

The outstanding combination of high resilience to mechanical deformations with rather strong strain-sensitive band structures makes two-dimensional (2D) semiconductors particularly suited for strain engineering [1–8]. These desirable properties have, indeed, triggered the interest of a great deal of the scientific community to research on the properties of 2D semiconductors under strain.

Molybdenum disulfide ($MoS_2$) is probably the most studied semiconductor to date [9–15] and several works focused on strain engineering [16–48]. Although many theoretical works predicted that biaxial strain can tune more effectively the band structure of $MoS_2$ [16–18,24,25,29,30,34,44], most of the experimental works only deal with the specific case of uniaxial strain [19,20,22,26–28,33,35,37–41,45–47,49,50]. To date, only a handful of experimental works explored the application of biaxial strain to atomically thin $MoS_2$ using piezoelectric substrates [21], thermal expansion mismatch [31,36,42,43,51], exploiting the presence of naturally occurring bubbles [48,52,53], the creation of artificial blisters [32,54,55] or bubbles [56–58], a thin film stressor method [59] or a capillary-pressure-induced nanoindentation method [60]. All these methods present some disadvantages (complexity, cross-talk, etc.) with respect to the beam-bending approach widespread to apply uniaxial strain, explaining the large number of works focused on uniaxial strain. Therefore, an experimental method that allows to control biaxial strain with a geometry similar to the beam bending method would be highly desirable. In 2015, Androulidakis *et al*. [61] adapted the macroscopic cruciform biaxial strain testing, used to probe the mechanical properties of standard materials, to apply biaxial strain to graphene. The method was based on the bending of a polymer substrate with cruciform shape through an indentation at its center. They applied this method to study the shift of Raman modes of graphene upon biaxial straining, but it has been overlooked by the community interested on strain engineering of 2D semiconductors (see the note after the conclusions).

Here we implement a simple experimental setup to apply biaxial strain to 2D materials, following the cruciform bending/indentation method, under the inspection of an optical microscope. We provide all the technical details to facilitate the replication of the setup by others, note that this relevant information was somewhat missing in Ref. [61] making it difficult adopting this technique by other experimental groups. We also modified a method recently developed to calibrate uniaxial straining setups [47] to calibrate the amount of biaxial strain achieved upon central indentation in the cruciform. We found that the calibration may strongly differ, depending on the specific dimensions of the cruciform, from the analytical formula used in Ref. [61] thus illustrating the relevance of performing an independent strain calibration. We then use the setup to strain 12 single-layer $MoS_2$ flakes finding that their reflectance spectra are red-shifted at a rate of -36 to -108 meV/% of biaxial tension. Interestingly, during the measurements we found that $MoS_2$ flakes are more prone to break during biaxial tensioning than during uniaxial tensioning (where the main failure mechanism is slippage). In many cases the breaking is followed by a sudden release of strain and further tensioning leads to a new red-shift of the reflectance spectra from the released position. We finally directly compare experimentally biaxial and uniaxial approaches by subjecting the same single-layer $MoS_2$ flake to successive biaxial and uniaxial tensioning cycles while monitoring the strain induced shift in the flake reflectance spectra. We experimentally verify that biaxial strain provides a more efficient way to tune the optical properties of $MoS_2$, as compared with uniaxial strain, in good agreement with theoretical predictions.

Figures 1(a-d) show pictures of the experimental setup develop to controllably bend cruciform polymer substrates through a central indentation. The setup is based on a manual linear Z-stage positioner (MAZ-40-10, by Optics Focus) that allows accurate displacements in the vertical direction (the minimum division of the micrometer screw correspond to 10 µm displacement). Figures 1(c-d) show an optical picture of a cruciform sample made of 100 µm thick Mylar placed onto the setup. Mylar is selected as substrate for the cruciform given its large Young's modulus (~4-5 GPa) as large strain transfer is obtained for substrates with a high Young's modulus [28,36] and it has been probed that a good strain transfer is already observed in substrates with a Young's modulus higher than 1 GPa [36,62]. Figures 1(e-p) show the blueprints of the homebuilt parts employed to modify the linear manual stage. The blueprint of the bracket is shown in Figures 1(e-h), it is the main part of the setup and the geometrical center of the flexible cruciform must be placed onto the center of its circular hole as illustrated in Figures 1(c-d). Figures 1(i-l) show the blueprint of the indenter, the hemisphere (Thorlabs PKFESP) placed on the top of this piece pushes the cruciform from the bottom and, therefore, it is symmetrically deformed in-plane [61]. The clamp, whose blueprint is shown in Figures 1(m-p), is responsible for holding the arms of the cruciform and let them slide over it while the geometrical center of the cruciform is being pushed by the indenter.

In order to directly calibrate the amount of biaxial strain that can be applied, for a given central indentation of the cruciform, we adapted the calibration protocol developed to calibrate uniaxial straining setups [47]. Briefly, we pattern an array of pillars with photolithography on the central part of the cruciform (Figure 2a) and we acquire optical microscopy images of the pillar array at different displacements of the micrometer screw. The biaxial strain value for a given micrometer screw displacement can be determined by measuring the distance between the pillars from the optical images as the strain, it is defined as:

$$\varepsilon = \frac{L - L_0}{L_0}$$

where $L_0$ is the pillar distance at zero-strain and $L$ at the given micrometer screw displacement.

Figure 2b shows the resulting biaxial strain calibration traces measured for 3 different polymer substrates: polycarbonate (PC, 250 μm), polypropylene (PP, 185 μm) and Mylar (100 μm). Moreover, by extracting the position of each pillar in the images (Figure 2c) one can even determine the spatial homogeneity of the applied biaxial strain and obtain a map of the spatial variation of the applied biaxial strain (Figure 2d), within a 500 by 500 μm$^2$ area around the center of the cruciform, finding a small variability of (2.1 ± 0.2) % strain (histogram reported in the Supporting Information).

To fabricate the single-layer MoS$_2$ samples to be studied, a bulk MoS$_2$ crystal (Molly Hill mine, Quebec, Canada) is exfoliated with Nitto tape (SPV224) and the cleaved crystallites are then transferred onto a Gel-Film substrate (WF x 4 6.0 mil, by Gel-Pak®). Single-layer flakes are identified on the surface of the Gel-Film substrate by combination of quantitative analysis of transmission mode optical microscopy images [63,64] and micro-reflectance spectroscopy [65,66]. Once a suitable single-layer MoS$_2$ flake is identified, it is deterministically placed onto the geometrical center of a cruciform within ~10 μm accuracy through an all dry transfer method [67–69]. The inset in Figure 3a shows a picture of a single-layer MoS$_2$ flake on a Mylar cruciform. We use differential micro-reflectance spectroscopy to probe the band structure changes induced by biaxial-strain on the single layer $MoS_2$ flake [66] (see Figure 3a). The spectra have two prominent peak features arising from the resonances associated to the direct valence-to-conduction band transitions at the K point of the Brillouin zone that yields the generation of excitons (labelled A and B according to the most common nomenclature in the literature) [10,65,70–72]. Upon biaxial tension, both A and B peaks red shift. Figure 3b shows the energy of the A and B peaks upon increasing biaxial strain. One can fit the excitons energy *vs*. strain dataset to a straight-line from whose slope the gauge factor, *i.e*. the excitons energy shift per % of biaxial tension, can be extracted. For the flake shown in Figure 3a we find gauge factor values of -90.2 meV/% and -81.5 meV/% for the A and B excitons respectively. The insets in Figure 3(b) show the statistical information obtained after measuring 12 different single-layer $MoS_2$ flakes. In these box-plots the dispersion of the obtained gauge factor can be observed. The box includes the data between the 25th and the 75th percentile, the middle line and small dot correspond to the median and the mean of the data, respectively, and the top and bottom lines correspond to the maximum and the minimum values, respectively. For A and B peaks, we found maximum gauge factor values of 108 meV/% and 102 meV/%, respectively. Table 1 shows a summary of the reported experimental gauge factors for biaxially strained $MoS_2$ in the literature, as a comparison. One can see how the gauge factor obtained through this cruciform bending method is significantly larger to that obtained through exploiting the thermal expansion of polypropylene substrates, pointing out that the strain transfer on polypropylene could be lower than the calculated values (close to 100%) or might be temperature dependent.

It is worth mentioning that we found that single-layer $MoS_2$ flakes are prone to break upon biaxial strain tension, and that the breakdown comes together with a sudden release of strain. Moreover, after cracking, one can typically keep strain-tuning the flake from the strain released energy position. This is in striking contrast to our previous observations in uniaxially strained TMDS flakes where the main source of failure during the straining tests was slippage that shows up as a drastic reduction of the strain gauge factor and hysteresis in the straining/releasing cycles. In Figure 4, a single-layer $MoS_2$ flake is biaxially strained at high strain values. Figure 4a shows the energy of the A peak while strain is increasing. At 1% strain, the flake cracks and the strain releases. After that, one can continue increasing the strain observing a new redshift of the excitons, with a different gauge factor, from the relaxed strain position. The bottom inset in Figure 4a presents the statistical information about the number observed flakes that crack upon biaxial strain. Six flakes break at certain strain, while 3 slip without breaking and another one slip first and then break. Two more flakes, subjected to a maximum strain of ~0.4% and 0.6% respectively, did not slip nor break. The top inset in Figure 4a also shows the distribution of critical strain values for cracking observed in the 7 single-layer $MoS_2$ fakes that cracked upon straining where a mean strain value around 0.74% leads to the breakdown of the flakes. Figures 4b and 4c show the flake before and after cracking. The red arrows point to the cracks that appeared in the flake right after observing the strain release in Figure 4a. Note that the biaxial strain induced shift of the excitons is reversible for strain loads below the slippage and failure strains. We address the reader to the Supp. Info. Figure S2 for a dataset of a single-layer $MoS_2$ flake subjected to several strain loading/releasing cycles.

Finally, we have directly compared the effect of biaxial and uniaxial strain to tune the micro-reflectance spectra on the same single-layer $MoS_2$. We first measured a biaxial strain cycle on a single-layer $MoS_2$ flake transferred onto the center of a Mylar cruciform (Figure 5a), similarly to Figure 3. After the measurement, two of the cruciform arms are cut away, transforming the sample into a simple beam, as shown in Figure 5b. We can then use a three-point bending test system [47] to subject the same single-layer $MoS_2$ to a uniaxial strain cycle. Figure 5c shows the strain dependent energy of the A and B excitons measured on the same flake subjected to a biaxial tensioning (0.6%) and uniaxial tensioning cycle (0.8%). While the gauge factor for the biaxial straining measurements is ~70 meV/%, for the uniaxial strain case it only reaches ~30 meV/% (in good agreement with our recent statistical analysis based on 15 single-layer $MoS_2$ flakes subjected to uniaxial strain [47]). This improved strain tunability for biaxial strain is attributed to be due to the fact that biaxial tension increases the interatomic distance in all in-plane directions while uniaxial strain, due to the Poisson's effect, increases the interatomic distance in the loading direction while compressing the lattice in the in-plane orthogonal direction. This orthogonal compression upon uniaxial loading effectively reduces the gauge factor. This experiment is, to our knowledge, the first experimental validation of the stronger effect of biaxial strain, as compared to uniaxial strain, to tune the band structure of $MoS_2$, predicted by DFT calculations [16–18,24,25,29,30,34,44].

CONCLUSIONS

In summary, we present all the details to implement a simple experimental setup to subject 2D materials to biaxial strain and we describe a protocol to accurately calibrate the amount of applied biaxial strain. We have applied the setup to study the strain-induced changes in the differential reflectance spectra of 12 single-layer $MoS_2$ flakes, finding strain-induced spectral redshifts with gauge factors ranging from 35 meV/% to 110 meV/%. Interestingly, we found that large biaxial strain tends to break single-layer $MoS_2$ (while slippage is more common on uniaxial straining experiments), thus suddenly releasing the accumulated strain. We also directly compare the strain tuning effectivity of biaxial and uniaxial strain by subjecting the same single-layer $MoS_2$ flake to biaxial and uniaxial strain cycles. This measurement experimentally validates previous theoretical predictions as we find a biaxial strain gauge factor 2.3 times the uniaxial strain one. We believe that the results shown here can help the community working on strain engineering of 2D materials to employ more and more biaxial strain and thus to achieve higher strain-induced band structure tunability.

**NOTE:** During the elaboration of this manuscript we became aware of the work of Michail *et al*. [73] where they use the cruciform bending/indentation method developed by Androulidakis *et al*. [61], similar to this work, to study the effect of biaxial strain in the photoluminescence and Raman spectra of exfoliated and chemical vapour deposited single- and bi-layer $MoS_2$. In our work we provide complementary information, not present in Ref. [73]: 1) details about the experimental setup, 2) details about the calibration of the biaxial strain, 3) micro-reflectance measurements, 4) statistical analysis of the biaxial strain gauge-factor, 5) analysis of the strain-induced failure of the devices and 6) direct comparison between uniaxial and biaxial strain tuning.

**Acknowledgements**

This project has received funding from the European Research Council (ERC) under the European Union's Horizon 2020 research and innovation programme (grant agreement n° 755655, ERC-StG 2017 project 2D-TOPSENSE). R.F. acknowledges the support from information of the breaking/slipping point, extracted from the 12 single-layer $MoS_2$ measured flakes. (b) Optical microscopy image of the $MoS_2$ flake before cracking. (c) Optical image of the $MoS_2$ after cracking.

FIGURES:

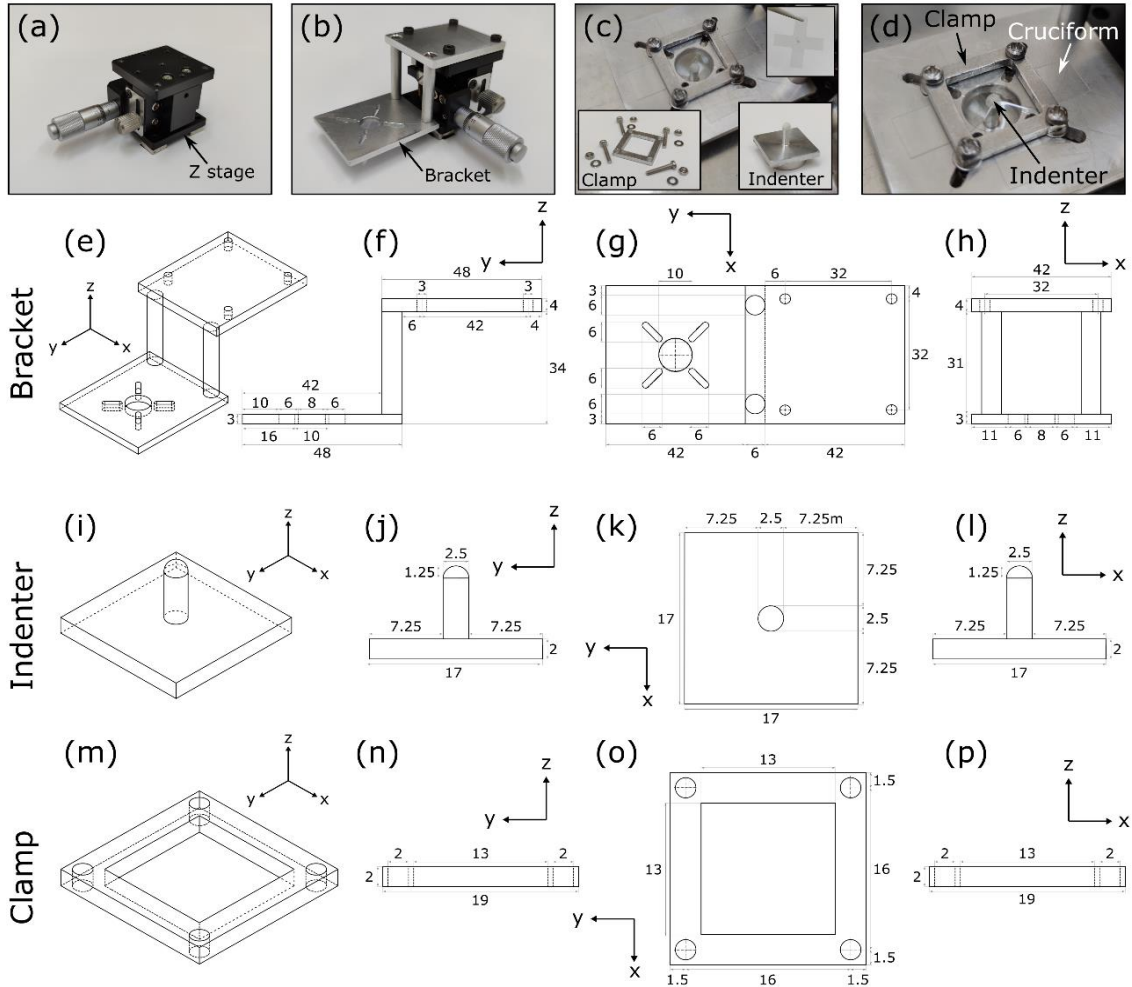

**Figure 1. Experimental setup for the bending/indentation of a cruciform polymer substrate for biaxial straining 2D materials.** (a-d) Pictures of the experimental setup. (e-h) Blueprint of the bracket. (i-l) Blueprint of the indenter. (m-p) Blueprint of the clamp. (All dimensions in millimeters).

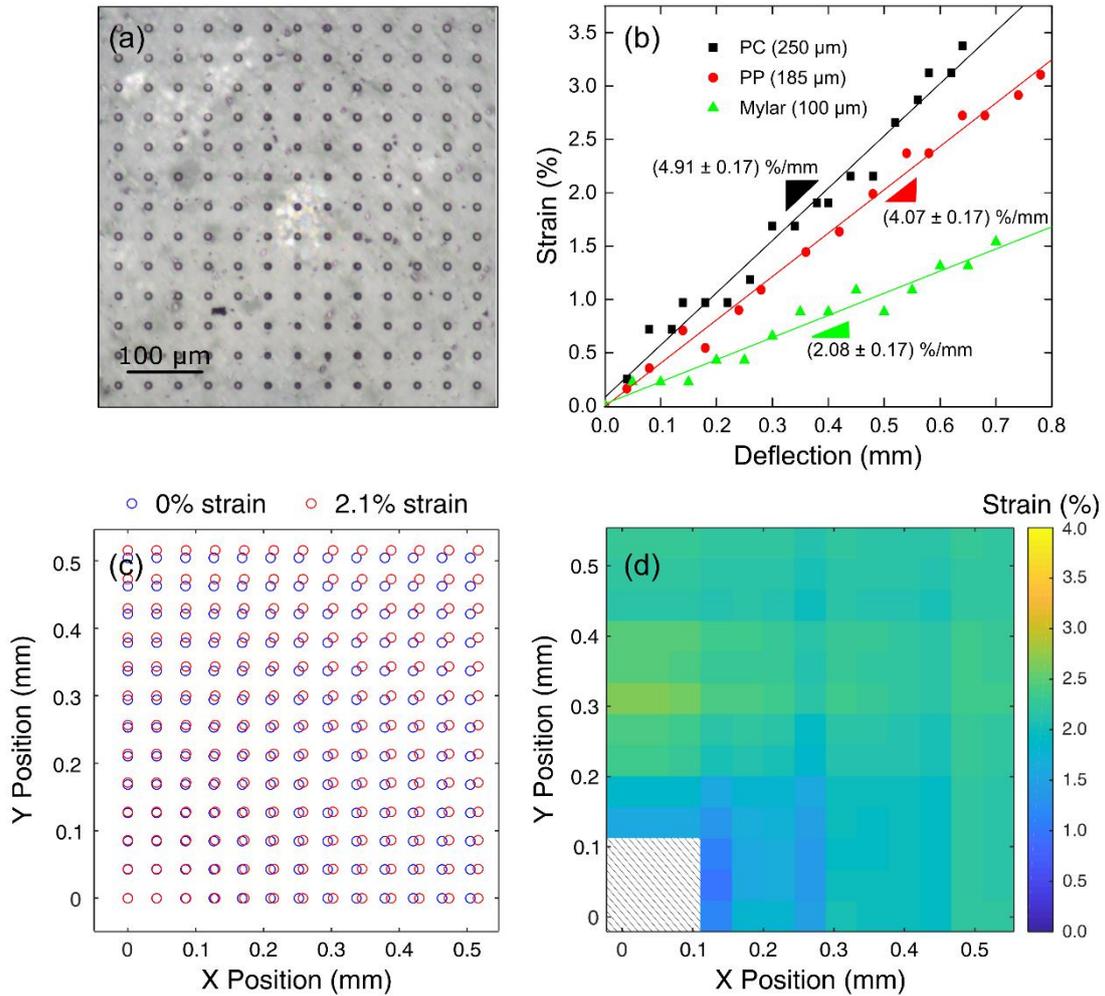

**Figure 2. Direct calibration of the applied biaxial strain.** (a) Optical picture of the patterned pillars. (b) Biaxial strain calibration for different polymer substrates. (c) extracted position of the pillars before and after applying strain. (d) Map of the spatial variation of the applied strain.

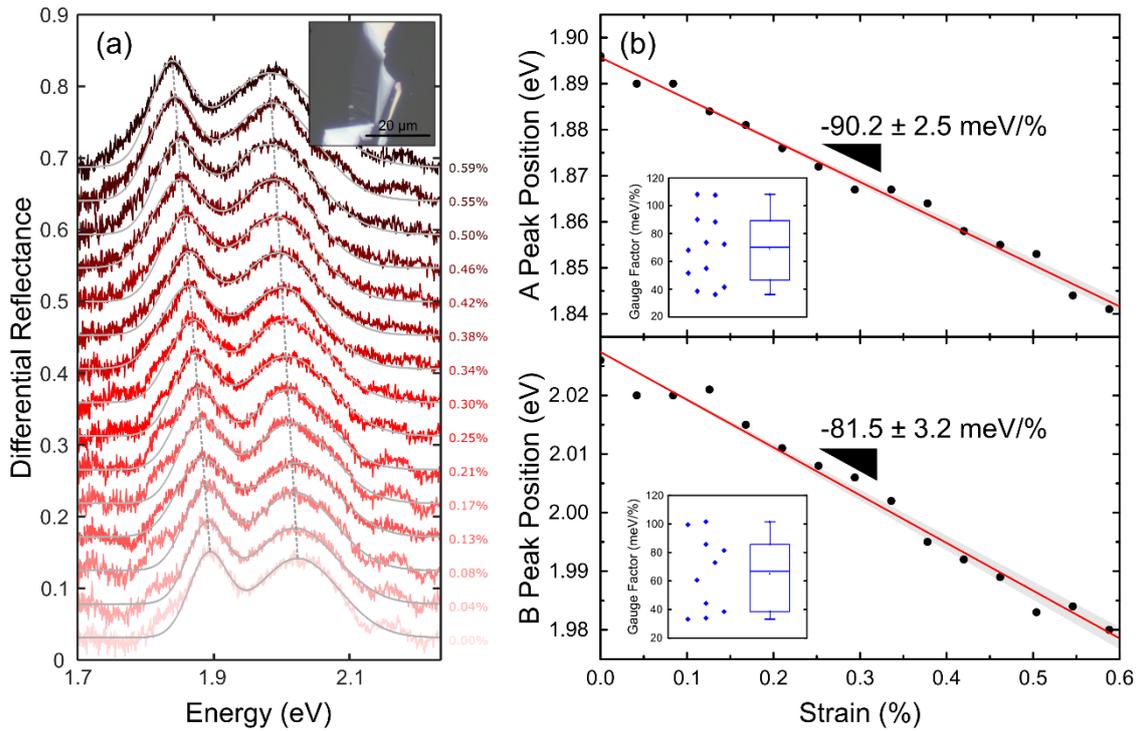

**Figure 3. Biaxial strain tuning the optical spectra of single-layer MoS$_2$.** (a) Differential reflectance spectra at different biaxial strain values of a single layer MoS$_2$ flake. (Inset) Optical microscopy image of the single-layer MoS$_2$ flake subjected to biaxial straion. (b) A and B exciton energy positions as a function of biaxial strain. Insets show the statistical information of the gauge factors obtained for 12 different single-layer MoS$_2$ flakes.

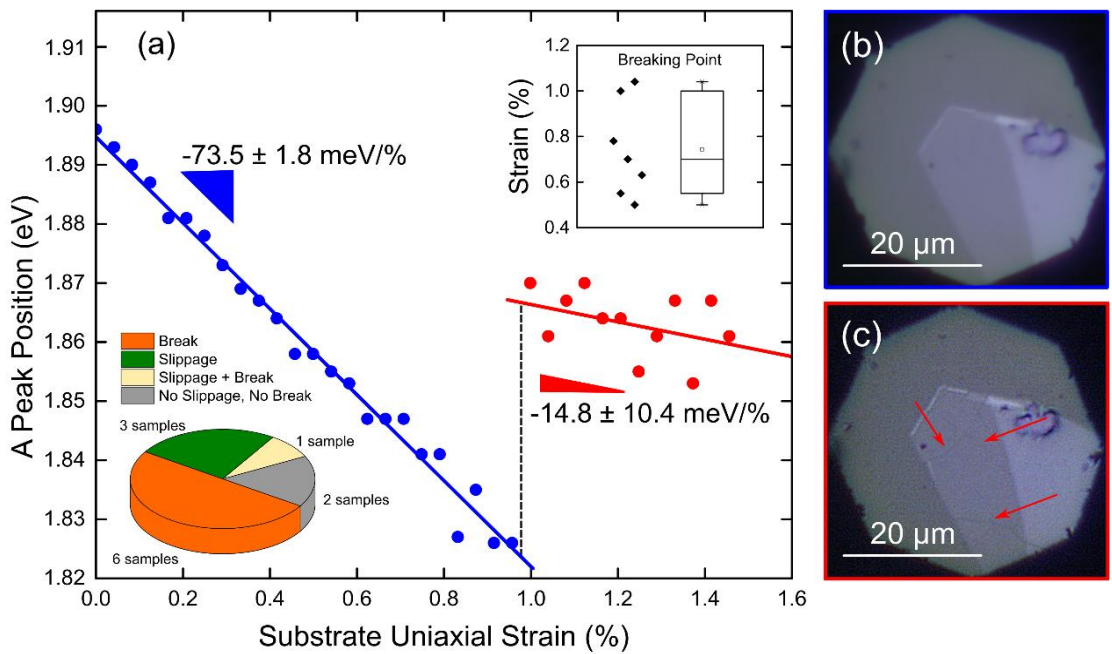

**Figure 4. Breakdown of single-layer MoS$_2$ flakes upon large biaxial strains.** (a) A exciton energy values as a function of biaxial strain of a single-layer MoS$_2$ flake. At 1% one can observe strain releases. Inset shows the statistical

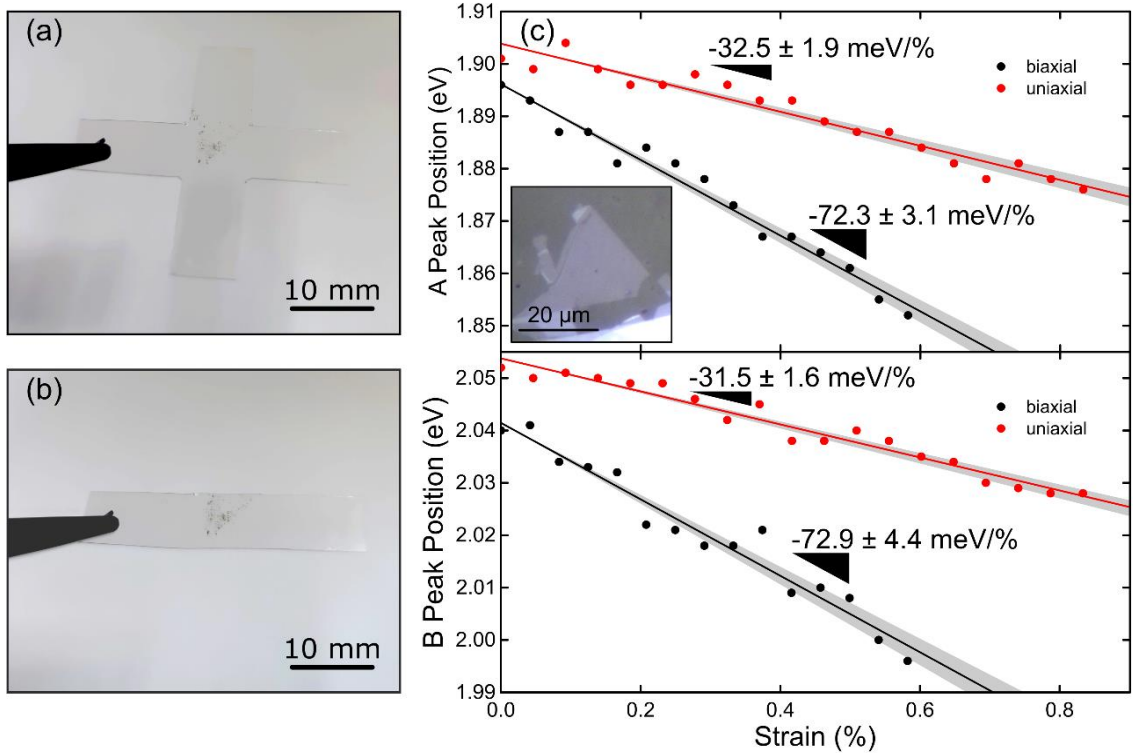

**Figure 5. Subjecting the same $MoS_2$ flake to biaxial and uniaxial strain.** (a) Picture of a cruciform with transferred single-layer $MoS_2$ flake use to test a biaxial strain experiment. (b) Same sample after cutting two of its arms in order test uniaxial strain. (c) A and B exciton energy positions measured on the same single-layer $MoS_2$ flake at different biaxial and uniaxial strain values. (Inset) Optical microscopy image of the single-layer $MoS_2$ flake under study.

| Work | Strain Method | Measurement | Number of layers | Substrate | Maximum strain (%) | A exciton gauge factor (meV/%) |
|---|---|---|---|---|---|---|
| **This work** | Bending/indentation cruciform | Micro-reflectance | 1L | Mylar | 1.04 | 108 |
| Hui et al. [21] | Piezoelectric substrates (compressive strain) | Photoluminescence and Raman | 3L | PMN-PT (piezoelectric substrate) | 0.2 | 300 |
| Plechinger et al. [31] | Thermal expansion mismatch | Photoluminescence | 1L | PDMS | 0.2 | 4.2* |
| Frisenda et al. [36] | Thermal expansion mismatch | Micro-reflectance | 1L | Polypropylene | 1 | 51.1 |
| Carrascoso et al. [43] | Thermal expansion mismatch | Micro-reflectance | 2L | Polypropylene | 0.87 | 41 |
| Gant et al. [42] | Thermal expansion mismatch | Photocurrent spectroscopy | 1L | Polycarbonate | -1.5 to 0.5 | 94 |
| Kyoung Ryu et al. | Thermal expansion mismatch | Micro-reflectance | 1L, 2L, 3L | Polypropylene | 0.64 | 1L: 48 2L: 55 3L: 32 |
| Guo et al. [48] | Naturally occurring bubbles | Photoluminescence and Raman | 2L, 3L, 4L, 5L | $SiO_2$/Si | 1 (in-plane strain) | 2L, 3L, 5L: 107 4L: 114 |
| Tyurnina et al. [52] | Naturally occurring bubbles | Photoluminescence | 1L | $MoS_2$ | 2 | 55 |
| Lloyd et al. [32] | Creation of artificial blisters | Photoluminescence | 1L | $SiO_2$/Si | 5.6 | 99 |
| Blundo et al. [56] | Creation of artificial bubbles | Photoluminescence | 1L | $SiO_2$/Si | 2.1 (radial strain) 4.2 (in-plain strain) | 37 |
| Yang et al. [57] | Creation of artificial bubbles | Photoluminescence and Raman | 1L, 2L, 3L | PDMS | 9.4 | 1L: 41 2L: 27.3 3L: 30 |
| Li et al. [60] | Capillary-pressure-induced nanoindentation | Photoluminescence and Raman | 1L | $SiO_2$/Si | 3 | 110 |
| Michail et al. [73] | Bending/indentation cruciform | Photoluminescence and Raman | 1L, 2L | PMMA | 0.88 | 124 (1L exfoliated) 76 (1L CVD) |

Table 1. Summary of the experimental gauge factors obtained for biaxially strained $MoS_2$ in literature. * In the original work a gauge factor of 105 mev/% was estimated but it did not account for the intrinsic thermal contribution to the redshift. After discounting this intrinsic thermal contribution, the gauge factor is 4.2 meV/%.

# Supporting Information:

# Biaxial versus uniaxial strain tuning of single-layer MoS$_2$

*Felix Carrascoso[1], Riccardo Frisenda[1] (\*), Andres Castellanos-Gomez[1] (\*)*

[1]*Materials Science Factory. Instituto de Ciencia de Materiales de Madrid (ICMM-CSIC), 28049, Madrid, Spain.*

Riccardo.frisenda@csic.es, Andres.castellanos@csic.es

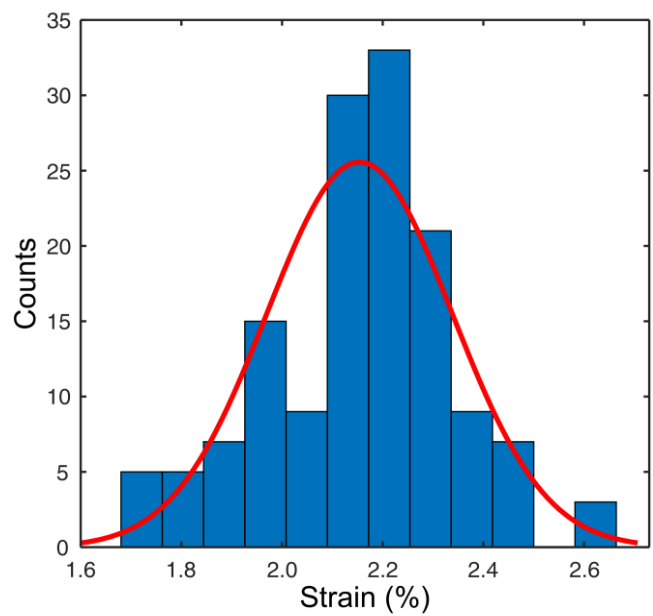

**Figure S1.** Histogram obtained from the Map of the spatial variation of the applied strain in Figure 2d.

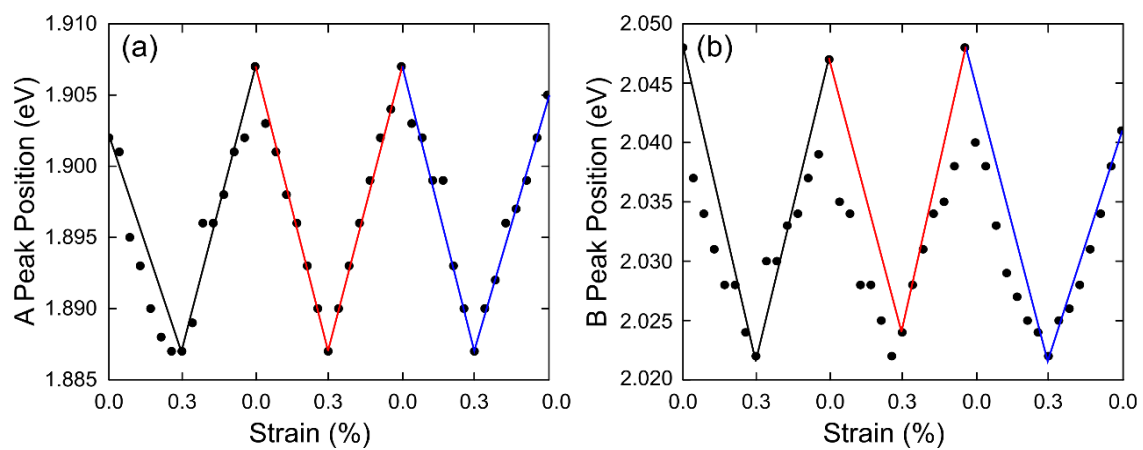

**Figure S2.** A and B exciton peak position energies as a function of biaxial strain during 3 strain loading/unloading cycles to illustrate the reversibility of the straining process when the strain level is below the slippage or break-down strain threshold.

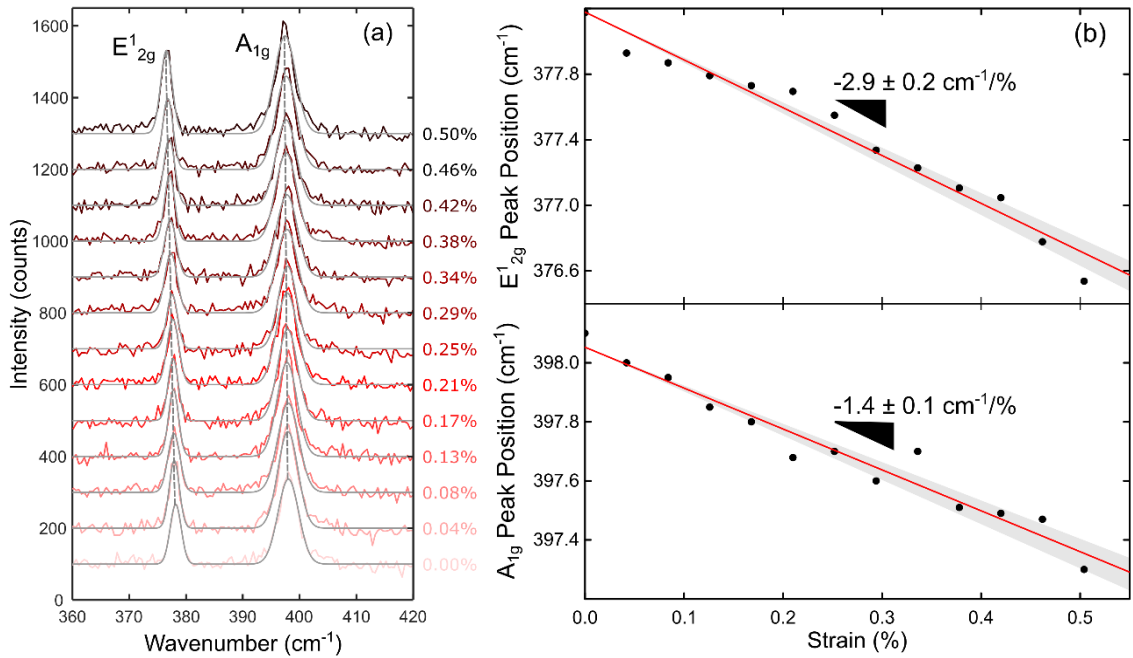

**Figure S3.** (a) Raman spectra acquired for different biaxial strains on a single-layer MoS$_2$ flake. The spectra have been shifted vertically by 100 counts for clarity. (b) E$^1_{2g}$ and A$_{1g}$ peak positions as a function of biaxial strain. A linear fit is used to determine the strain tuning Raman shift rate.

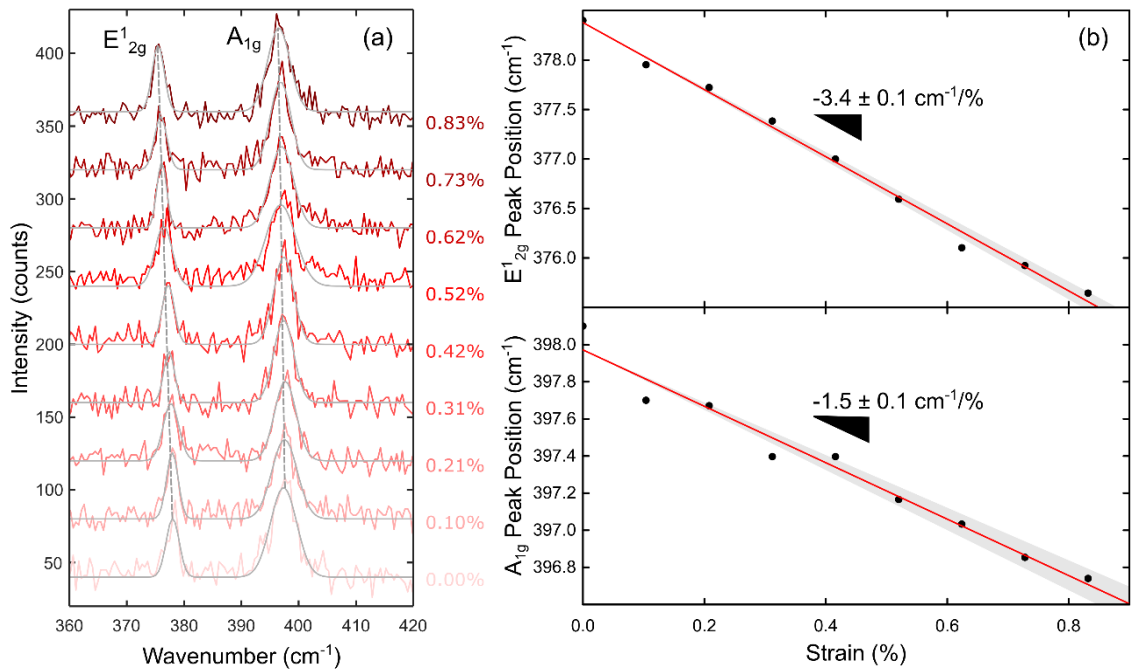

**Figure S4.** (a) Raman spectra acquired for different biaxial strains on a single-layer MoS$_2$ flake. The spectra have been shifted vertically by 40 counts for clarity. (b) E$^1_{2g}$ and A$_{1g}$ peak positions as a function of biaxial strain. A linear fit is used to determine the strain tuning Raman shift rate.

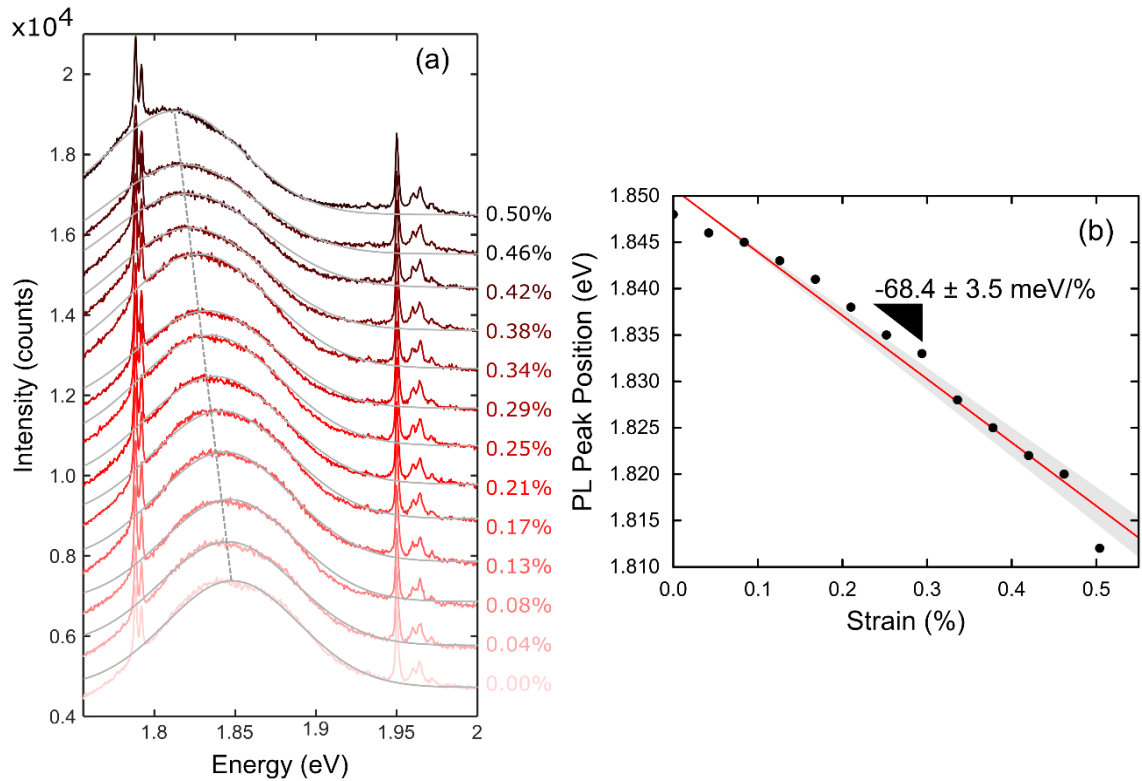

**Figure S5.** (a) Photoluminescence spectra acquired for different biaxial strains on a single-layer $MoS_2$ flake. The spectra have been shifted vertically by 1000 counts for clarity. (b) A exciton photoluminescence peak positions as a function of biaxial strain. A linear fit is used to determine the biaxial strain gauge factor.

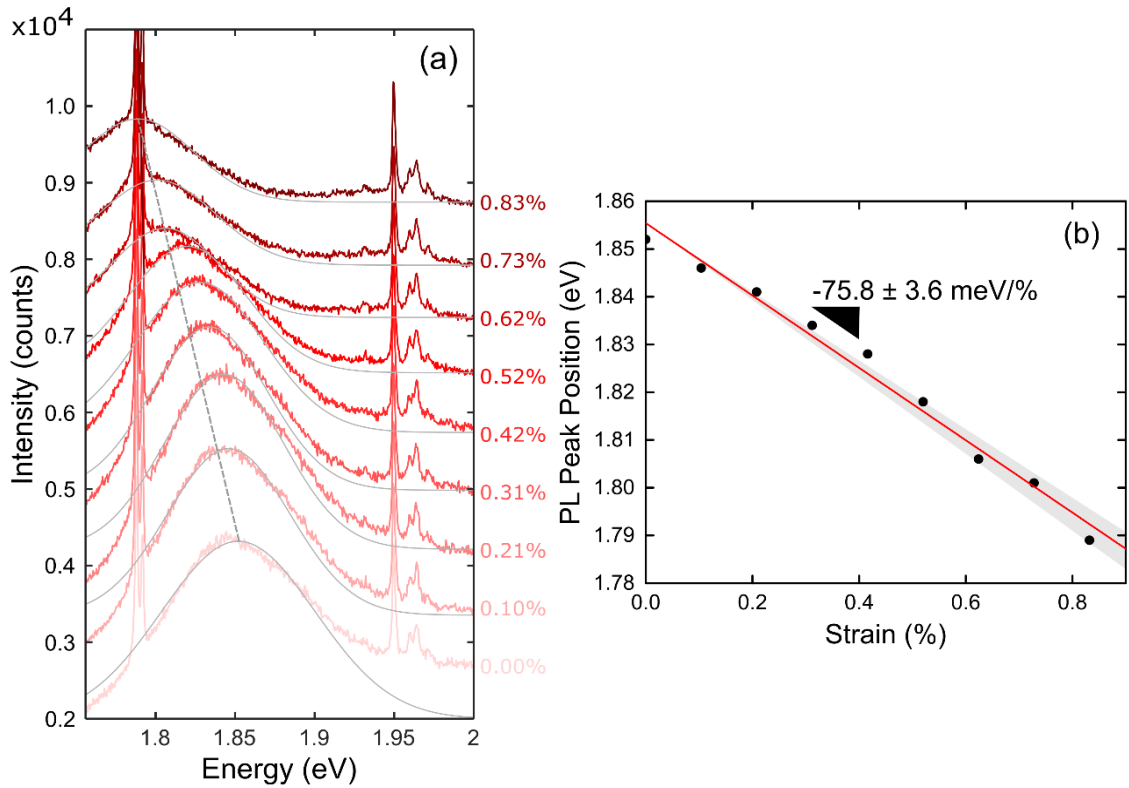

**Figure S6.** (a) Photoluminescence spectra acquired for different biaxial strains on a single-layer MoS$_2$ flake. The spectra have been shifted vertically by 700 counts for clarity. (b) A exciton photoluminescence peak positions as a function of biaxial strain. A linear fit is used to determine the biaxial strain gauge factor.